\documentclass[aps,prb,twocolumn,superscriptaddress]{revtex4-2}
\usepackage{makeidx}
\usepackage{graphicx}
\usepackage{amsmath}
\usepackage{amsfonts}
\usepackage{amssymb}
\usepackage{bm}
\usepackage{xcolor}
\usepackage{float}
\usepackage{physics}
\usepackage{hyperref}

\begin{document}
\title{Symmetry breaking and superfluid currents in a split-ring spinor polariton condensate}

\author{Igor Chestnov}
\email{igor\_chestnov@mail.ru}
\affiliation{School of Physics and Engineering, ITMO University, Saint Petersburg 197101, Russia}

\author{Kirill Kondratenko}
\affiliation{Moscow Institute of Physics and Technology, 141701 Institutskii per., Dolgoprudnyi, Moscow Region, Russia}
\affiliation{Russian Quantum Center, Skolkovo IC, Bolshoy Bulvar 30, bld. 1, Moscow 121205, Russia}

\author{Sevak Demirchyan}

\affiliation{Russian Quantum Center, Skolkovo IC, Bolshoy Bulvar 30, bld. 1, Moscow 121205, Russia}

\author{Alexey Kavokin}

\affiliation{Key Laboratory for Quantum Materials of Zhejiang Province, School of Science,
Westlake University, 18 Shilongshan Road, Hangzhou 310024, Zhejiang, China}
\affiliation{Institute of Natural Sciences, Westlake Institute for Advanced Study, 18 Shilongshan Road, Hangzhou,
Zhejiang Province 310024, China}
\affiliation{Spin Optics Laboratory, St. Petersburg State University, Ulyanovskaya 1, St. Petersburg 198504, Russia}

\date{\today}

\begin{abstract}
Bosonic condensates of spin-less non-interacting particles confined on a ring cannot propagate circular periodic currents once rotation symmetry of the system is broken. However a persistent current may appear due to inter-particle interactions exceeding some critical strength.  In this up-critical regime breaking of the symmetry between the clockwise and anticlockwise rotations takes place. We consider this symmetry-breaking scenario in the case of a spinor condensate of exciton polaritons trapped on a ring split by a potential barrier. Due to the intrinsic symmetry of the effective spin-orbit interaction which stems from the linear splitting between transverse-electric and transverse-magnetic microcavity modes, the potential barrier blocks the circulating current and imposes linear polarization patterns. On the other hand, circularly polarized polaritons form circular currents propagating in opposite directions with equal absolute values of angular momentum. In the presence of inter-particle interactions, the symmetry of clockwise and anticlockwise currents can be broken spontaneously. We describe several symmetry-breaking scenarios which imply either restoration of the global condensate rotation or the onset of the circular polarization in the symmetry-broken state.
\end{abstract}

\maketitle

\section{Introduction}

Quantum mechanics implies that a non-degenerate state of a single-particle system always possesses its intrinsic symmetry.
This rule, however, ceases to be strict in the many-body regime where nonlinear self-interactions give rise to the  states with reduced symmetry. 
This phenomenon known as \textit{spontaneous symmetry breaking} (SSB) \cite{Malomed2013Book} was observed in various physical contexts including nonlinear optics \cite{Jensen1982,Cao2017,delBino2017}, bosonic Josephson junctions \cite{Raghavan1999,Albiez2005}, metamaterials \cite{liu2014} etc.

A prototypical system featuring SSB is a bosonic condensate trapped in a symmetric double-well potential \cite{Albiez2005}. 
In the linear regime, the coherent tunneling across the barrier establishes symmetric occupation of two potential minima. 
However beyond the limit of weak inter-particle interactions, a macroscopic quantum self-trapping phenomenon occurs when nonlinear effects dominate coherent coupling. 
It violates an underlying spatial-parity symmetry and imposes steady-state population imbalance between the wells \cite{Abbarchi2013}. 

Self-trapping is characteristic of nonlinear systems composed by two linearly coupled components.
For example in the two-component or \textit{spinor} condensates, the SSB is manifested in breaking an \textit{internal parity} symmetry connected with the invariance under swapping of the system components \cite{Zibold2010,Ohadi2015}. 

Similarly, in the annular geometry, the SSB violates a chiral symmetry which implies equivalence of the clockwise (CW) and counterclockwise (CCW) rotations. 
In the presence of a potential scatterer which couples CW and CCW waves, the chiral symmetry supports standing wave solutions which carry no net circulation \cite{Kippenberg2002}. It requires time-reversal symmetry which guarantees reciprocal scattering between CW and CCW waves.
Reduction of the symmetry at the SSB event is manifested by the establishment of the global rotation as it was observed in bosonic condensates \cite{Wright2013,Askitopoulos2018} confined in the annular traps and in whispering-gallery microcavities \cite{Cao2017,delBino2017}.

Studying SSB in the nonsimply connected geometry such as a ring is motivated by the possibility of creating persistent current without stirring. 
Being a hallmark of superfluidity, persistent current has been mostly investigated in the conventional superfluids characterised by a scalar order parameter.
Extending this concept to the spinor case \cite{Beattie2013} naturally requires accounting for the interaction between internal and orbital degrees of freedom, either intrinsic \cite{Splettstoesser2003} or synthetic \cite{Dalibard2011}.

The interplay of nonlinearity and spin-orbit coupling (SOC) in a ring-shaped atomic condensates has been investigated in the presence of the artificial Rashba-type \cite{White2017} and combined Rashba-Dresselhaus \cite{Karabulut2016} SOC mechanisms. 
On the other hand, in photonics, SOC naturally appears at the sub-wavelength scale \cite{Bliokh2015} and in microcavities \cite{KavokinMicrocavities}. Besides, the presence of a birefringent media \cite{Rechcinska2019,ren2021} such as liquid crystal-filled microcavities allows for a flexible manipulation of the synthetic spin-orbit interaction for light.

In this manuscript, we consider a weakly linked annular spinor superfluid subject to the photonic analog of SOC which is inherently present in semiconductor microcavities. We refer to a split-ring geometry that has been recently put forward as promising for the realization of qubits based on superpositions of superfluid polariton currents \cite{Xue2021,Kavokin2022}.

The SOC in this system originates from the momentum-dependent splitting between linearly polarized transverse-electric (TE) and transverse-magnetic (TM) photonic modes \cite{Panzarini1999}. 
A strong coupling between the optical mode and the quantum-well excitons gives rise to the hybrid quasiparticles known as exciton polaritons. 
Being composite bosons, polaritons can form a macroscopic coherent state analogous to Bose-Einstein condensate \cite{kasprzak2006} with intrinsically two-component order parameter inherited from photon polarization. 
An important advantage of polaritons is their strong inter-particle interactions stemming from the exciton component which is a necessary ingredient of the nonlinear symmetry breaking.

In the system with broken rotation symmetry, excitation of the circular polariton current 
requires an explicit breaking of the chiral symmetry which can be realised by means of synthetic gauge field \cite{Shelykh2009,Chestnov2021}, gain asymmetry \cite{Sedov2021PRR,Yulin2020} or with the use of external driving \cite{Yao2023,Kwon2019}. 
Contrary to these approaches, the SSB triggered by polariton-polariton interactions allows for spinning ring-shaped condensate in the system which intrinsically respects chiral symmetry. 

The effect of spontaneous polariton rotation analogous to the chiral symmetry breaking has been predicted for the spinless condensate devoid of SOC \cite{Nalitov2019}.
The impact of TE-TM splitting in the ideal (defectless) polariton ring was analysed in \cite{Gulevich2016}. In this case, 
polariton-polariton interactions tend to break continuous rotational symmetry and favour formation of spatially nonuniform condensate.
In this paper, we focus on the interplay of the symmetry imposed by the TE-TM splitting and the potential responsible for the rotation symmetry breaking. 

The paper is organised as follows. In Sec.~\ref{Sec:LinearProblem} we investigate the linear eigenstates of the ring-shaped polariton condensate. Here we focus on the properties of the combined symmetry provided by the TE-TM splitting. The manifestations of the SSB triggered by the polariton-polariton interaction are discussed in Sec.~\ref{Sec:NonlinearSSB}. Treating the problem numerically, we reveal various types of symmetry-breaking solutions. Then adopting the simplified four-wave model which accounts for the two contra-propagating  vortices in each polarization \cite{yulin2022}  we analyze the most prominent SSB scenarios. The elaborated simplified description allows for the analytical prediction of the SSB threshold density.

\section{Spinor polariton condensate in a split-ring and its symmetry properties}\label{Sec:LinearProblem}

\subsection{The model}
\begin{figure}
    \centering
    \includegraphics[width=0.6\linewidth]{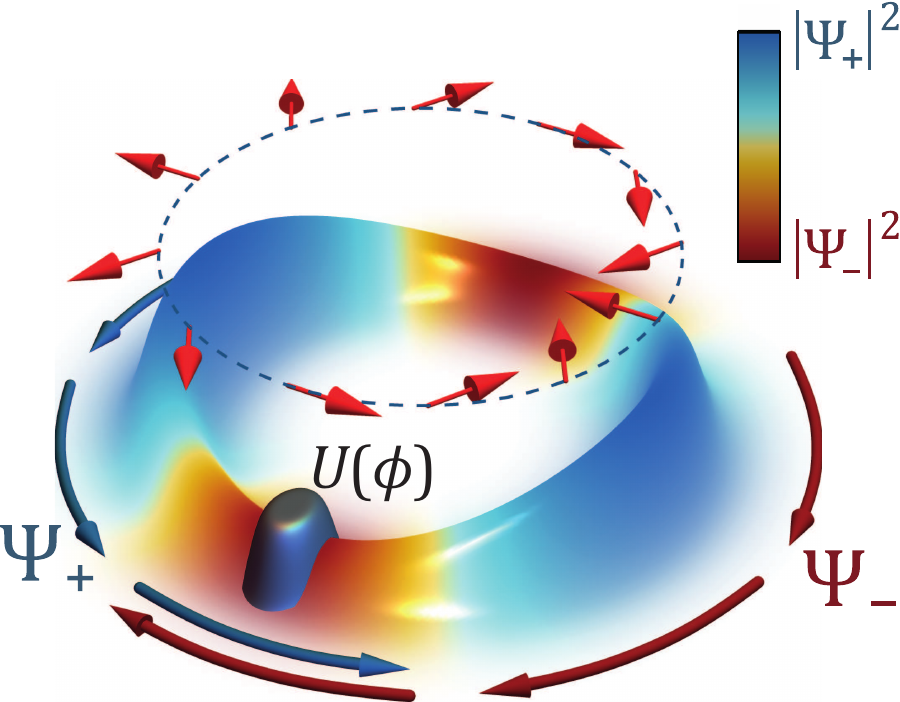}
    \caption{The ring-shaped polariton condensate split by the defect potential $U(\phi)$. 
    The blue and the vinous arrows indicate the presence of the persistent currents of circularly polarized polaritons. 
    Red straight arrows indicate the direction of the effective magnetic field associated with the TE-TM splitting. 
    The false color of the condensate density encodes a local value of the degree of circular polarization.
    }
    \label{Fig:Sketch}
\end{figure}

To be specific, we consider a spinor polariton condensate confined on a ring as sketched in Fig.~\ref{Fig:Sketch}. The annular trapping can be realised with the use of various approaches including sculpting the pump beam with spatial light modulator 
 \cite{Yao2023}, using pillar microcavities \cite{Kalevich2014} or etching of ring-channels \cite{Mukherjee2021}.
The order parameter of the condensate is characterised by the two-component complex spinor $\bm{\Psi} = \left(\Psi_{+},\Psi_{-}\right)^\intercal$ written in the basis of opposite circular polariton polarizations. 
In the mean-field treatment, $\bm{\Psi}$ obeys two-dimensional Gross-Pitaevskii equation \cite{KavokinMicrocavities}.  However a tight radial confinement characteristic of a thin ring allows to split $\bm{\Psi}$ into independent radial and azimuthal components.
Thus integrating out the radial dependence yields an effective one-dimensional Gross-Pitaevskii equation (GPE) \cite{Kozin2018,Mukherjee2021}: 
\begin{equation}\label{GPE}
  i\hbar \partial_t\bm{\Psi}= \left[\hat{\mathcal{H}}_{\rm 0} + \hat{\mathcal{H}}_{\rm int} \right] \bm{\Psi}.
\end{equation}
Here the first term stands for the single-particle Hamiltonian: 
\begin{equation} \label{SPHam}
    \hat{\mathcal{H}}_{\rm 0} = \frac{\hbar^2}{2m^\star} \begin{pmatrix}
    -\partial_{\phi\phi}^2 + U(\phi)     &   \Delta e^{-2i\phi}          \\
    \Delta e^{2i\phi}                   &   -\partial_{\phi\phi}^2 + U(\phi)
    \end{pmatrix},
\end{equation}
with  $m^\star$ being the effective mass characterising polariton azimuthal motion. 
In a thin ring, $m^\star$ can be connected with the 2D polariton effective mass by $m^\star = m_{\rm pol}R^2$. $\phi$ is the polar angle, $U(\phi)$ is the spin-independent potential. 

The effect of momentum-dependent splitting between linearly polarized TE and TM modes is described by the off-diagonal terms in Eq.~\eqref{SPHam}. 
Since the splitting is quadratic in the in-plane polariton momentum \cite{Kavokin2005}, the effect of SOC is governed by the second angular harmonics. 
In the 1D case, the SOC strength $\Delta$ is proportional to the inverse square of the ring width \cite{Kozin2018,Mukherjee2021} and the degree of asymmetry between effective masses of the TE and TM polariton modes. 
Thus tuning confinement potential or the microcavity properties, one can  vary $\Delta$ in a wide parameter range. 
In this paper, we focus on the case of weak TE-TM splitting, namely, at $\Delta < 1$. 
This limit allows us to approach the problem analytically.

The inter-particle interactions which are expected to be responsible for the symmetry breaking effect are described by the second term in Eq.~\eqref{GPE}:
\begin{equation*}
    \hat{\mathcal{H}}_{\rm int} = \frac{\hbar^2}{2m^\star} \begin{pmatrix} g|\Psi_{+}|^2 & 0 \\ 0 & g|\Psi_{-}|^2 \end{pmatrix},
\end{equation*}
which accounts for the repulsion between the same-spin polaritons with the strength $ g$. Here we neglect the interaction between the cross-polarized polaritons as it is typically weak \cite{Vladimirova2010}. 
Besides, in order to focus on the symmetry properties imposed by the SOC, we neglect the driven-dissipative effects. 
This approximation corresponds to the quasi-conservative limit realized in the ultrahigh-finesse microcavities \cite{Mukherjee2021}. 
Thus we consider a general model of a spinor condensate with the fixed total number of particles $N_{\rm pol} = \int \bm{\Psi}^{\dagger} \bm{\Psi} d\phi$. 
As a parameter controlling the strength of nonlinear effects we use the average polariton density, $\rho = N_{\rm pol}/2\pi$. 

In what follows, we are interested in the steady-state solutions:
\begin{equation}
\bm{\Psi}(t, \phi) = \bm{\Psi}(\phi) e^{-i\mu \frac{E_1}{\hbar} t}.
\end{equation}
For the sake of simplicity of notations, we operate with the dimensionless condensate energy $\mu$ which is given in the units of $E_1 = \hbar^2/2m^{\star}$. 

Since in the scalar case, SSB is manifested by the finite net circulation, we focus on quantifying rotation of the spinor condensate. In particular, we characterise it by the average orbital angular momentum (OAM) which can be defined for each pseudospin component individually,
 \begin{equation}\label{Eq_OAM}
 	\ell^{\pm} =\frac{m^\star } {\hbar N_{\rm pol}} \int j_{\pm}(\phi)\, d\phi,
 \end{equation}
where $j_{\pm}(\phi) = \hbar/{m^\star}\Im\left[ \Psi_\pm^*\partial_\phi\Psi_\pm  \right]$ is the density current of the same-spin polaritons. Definition \eqref{Eq_OAM} ensures additivity of the total angular momentum per particle,
\begin{equation}
    \ell^{\rm tot} = \ell^{+} + \ell^{-},
\end{equation}
which is the main parameter characterizing the net circulation of the spinor condensate.

\subsection{The symmetry imposed by the TE-TM splitting}

First, we focus on the symmetry properties of single-particle states which are governed mainly by the presence of SOC.
The symmetry imposed by the TE-TM splitting naturally follows from the time-reversal invariance of the two uncoupled linearly polarized microcavity modes. 
As it is shown in \cite{Rubo2022}, in the circular basis, this symmetry corresponds to the invariance under the transformation $\hat{\mathcal{T}} = \hat{\mathcal{K}} \hat{\sigma}_x$ where $\hat{\mathcal{K}}$ stands for the complex conjugation and $ \hat{\sigma}_x$ is a spin-flip operation which swaps circular polarizations. 
Note that in the spinless case, the chiral symmetry requires invariance under rotation inversion which is governed by the time-reversal operator $\hat{\mathcal{K}}$. 
For the spinor polariton condensate, this operation has to be supplemented by flipping of the effective spin $\hat{\sigma}_x$. 

Since the linear Hamiltonian $\hat{\mathcal{H}}_0$ commutes with $\hat{\mathcal{T}}$, the single-particle states of the condensate are expected to be either symmetric (even) or antisymmetric (odd) with respect to time reversal. 
The general form of these states reads:
\begin{equation}\label{TREigenStates}
    \bm{\Psi}_{\rm e,o}(\phi) = \begin{pmatrix} \Psi_0(\phi) \\ \pm \Psi_0^{*}(\phi) \end{pmatrix},
\end{equation}
where e(o) stands for even (odd) states and the wave function $\Psi_0(\phi)$ is governed by the eigenvalue problem for $\hat{\mathcal{H}}_0$. 
The states \eqref{TREigenStates} have two important properties. As long as $\left|\Psi_{+}\right| = \left| \Psi_{-} \right|$, the condensate is linearly polarized 
in \textit{any energy state}. Besides, if $\Psi_+$-component exhibits nonzero net circulation, the opposite circular polarization $\Psi_-$ rotates in the opposite direction with the same absolute value of OAM such as $\ell^{+} = - \ell^{-}$, see Eq.~\eqref{Eq_OAM}. Therefore, the total circulation vanishes, $\ell^{\rm tot} = 0$. Hence, these states are analogues to the so-called hidden-vortex states \cite{Brtka2010} predicted for the binary atomic condensates. 

Note that the discussed symmetry is violated if the time-reversal symmetry is broken explicitly, for example, in the presence of magnetic field \cite{Shelykh2009} which splits circular polarizations or due to the stirring potential \cite{gnusov2023,yulin2022} which breaks equivalence of the CW and CCW rotations. 
However, the effects of dissipation associated with the finite polariton lifetime do not violate chirality so that the symmetry properties of the condensate remain unaltered. 
In what follows, under the time-reversal symmetry breaking we imply the deviation of the condensate symmetry from the condition Eq.~\eqref{TREigenStates}. 

\subsection{Single-particle states}\label{SecSPstates}

To verify our symmetry analysis, we investigate the set of single-particle eigenstates of Hamiltonian \eqref{SPHam}. It is instructive to start from the rotation-symmetric case which has been exhaustively studied in \cite{Gulevich2016}. In particular, at $U=0$ the energy spectrum of the ring-shaped polariton condensate reads:
\begin{equation}\label{eq_TETM_empty_energies}
    \mu_{n\pm}= 1 + n^2 \pm \alpha,
\end{equation}
while the corresponding spinor wave functions can be written in the form \begin{equation}\label{eq_TETM_empty_eigstates}
     \mathbf{\Psi}_{n\pm} = \frac{\sqrt{N_{\rm pol}}}{\sqrt{4\pi\alpha}\sqrt{\alpha \pm 2n}}\begin{pmatrix} \Delta e^{-i\phi} \\ \pm \left(\alpha \pm 2n \right)e^{i\phi} \end{pmatrix} e^{i n \phi}.
\end{equation}
Here $n$ is integer and $\alpha = \sqrt{\Delta^2 + 4n^2}$. 

Since $\mu$ is quadratic in the solution index $n$, the energies of the states $n$ and $-n$ coincide. 
Because of this intrinsic degeneracy, the $n \neq 0$ solutions \eqref{eq_TETM_empty_eigstates} violate the symmetry of the time-reversal invariant states \eqref{TREigenStates}. 
Indeed, since any linear combination 
\begin{equation}\label{Eq_SuperPos}
\mathbf{\Phi}_{n\pm} = C_1 \mathbf{\Psi}_{n\pm} + C_2\mathbf{\Psi}_{-n\pm}
\end{equation}
is a suitable eigenmode, the total angular momentum $\ell^{\rm tot}_{n\pm}$ of the states $\mu_{n\pm}$ is not fixed but falls within the range from $-n(\alpha\pm 2)/\alpha$ to $n(\alpha\pm 2)/\alpha$. 
The only exception is the time-reversal eigenstates $\mathbf{\Psi}_{0\pm} = \sqrt{N_{\rm pol}/4\pi}\left(e^{-i\phi}, \pm e^{i\phi}\right)^\intercal$ which are linearly polarized in tangential and radial directions, respectively, and have zero total OAM \cite{Rubo2022}.

\begin{figure}
    \centering
    \includegraphics[width = \linewidth]{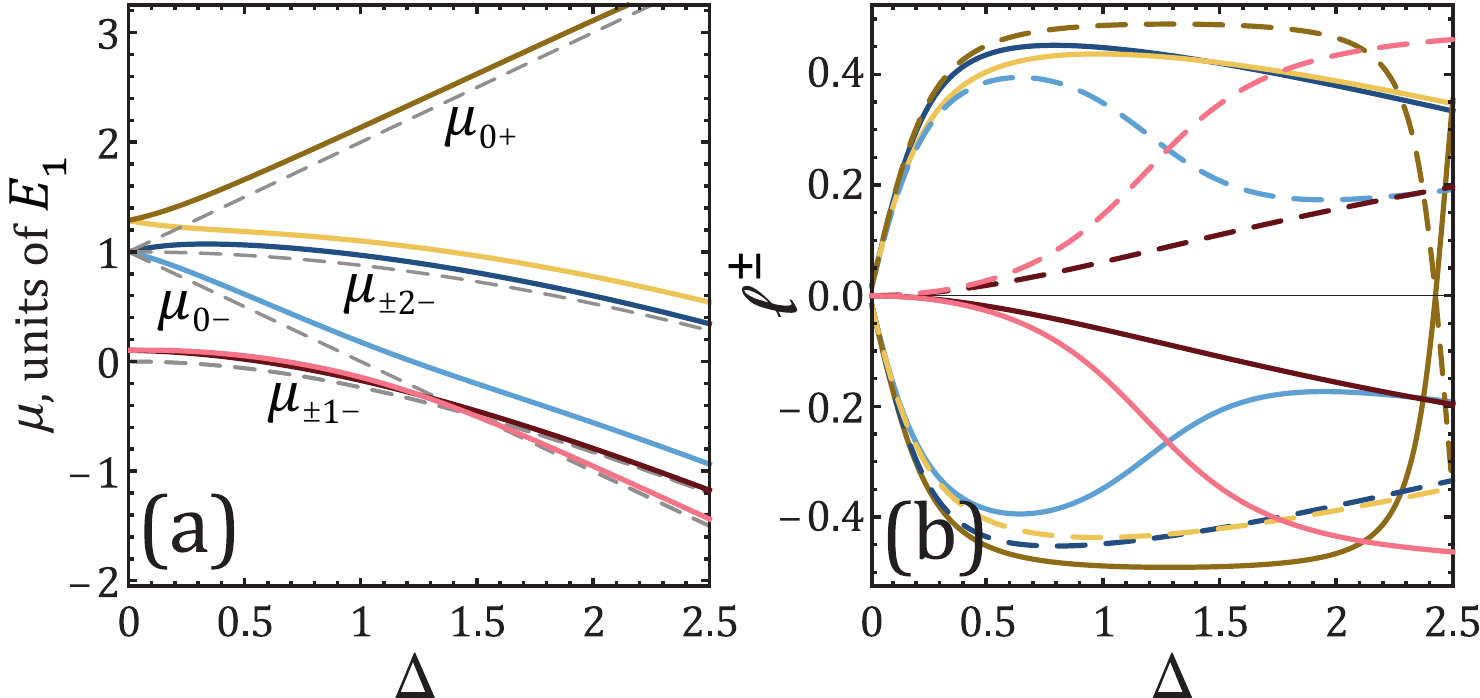}
    \caption{Non-interacting polariton condensate confined on a ring. 
    (a) The energy spectrum in the presence of the point-like defect $U(\phi) = 2 \pi V_0 \delta(\phi)$ (solid lines) at $V_0 = (2 \pi)^{-1}$. 
    The energy levels  \eqref{eq_TETM_empty_energies} of the defectless ring are shown by grey dashed lines. 
    The scale on the horizontal axis corresponds to the energy of the scalar vortex state with $2\pi$ phase winding. 
    (b) The average OAM of the circularly polarized polaritons $\ell^{\pm}$. The colors are consistent with those in (a). 
    The solid lines correspond to the OAM of the left-circularly polarized polaritons $\ell^+$, while the dashed lines describe right-circularly polarized polaritons, $\ell^{-}$. }
    \label{fig1}
\end{figure}

The considered degeneracy originates from the continuous rotational symmetry. 
Therefore, it gets lifted by any external potential so that the resulted eigenstates acquire $\hat{\mathcal{T}}$-symmetric properties \eqref{TREigenStates} provided $U(\phi)$ is real-valued and thus invariant under time reversal.
As a practically important example, we consider a point-like defect potential $U(\phi) = 2 \pi V_0\delta(\phi)$ which models a structural impurity of the microcavity sample. The corresponding energy splitting is demonstrated in the $\Delta$-dependence of the energy spectrum, Fig.~\ref{fig1}a.

As in the spinless problem \cite{Kippenberg2002},
the defect prohibits net circulation of the spinor condensate, $\ell^{\rm tot} = 0$. 
However, because of the presence of the TE-TM splitting, equal fractions of circularly polarized polaritons spin in opposite directions. 
The resulted spin textures exhibit linear polarization at any angular position on the ring that is protected by the time-reversal symmetry of the problem.

The corresponding $\ell^{\pm}(\Delta)$-dependencies are shown in Fig.~\ref{fig1}b.  
At the weak TE-TM splitting, the potential tends to suppress circulations of the individual spin components. 
In the opposite limit of the dominating SOC, $\Delta/V_0 \gg 1$, the states which stem from the splitting of the defect-free solution $\mathbf{\Psi}_{n\pm}$ can be represented as symmetric ($C_1 = C_2$) and anti-symmetric ($C_1 = - C_2$)  combinations \eqref{Eq_SuperPos} with
\begin{equation}\label{Eq.TETM_LinearCurrents}
\ell^{+}_{n\pm} = - \ell^{-}_{n\pm} = -\frac{1}{2}\left( 1 \pm \frac{2n^2}{\alpha} \right).
\end{equation}
Therefore, in this regime, the difference of OAMs of circularly polarized polaritons $\Delta \ell = \ell^{-} - \ell^{+}$ approach $1$ in any state.

\section{Spontaneous breaking of the time reversal symmetry}\label{Sec:NonlinearSSB}

In this section we investigate the impact of self-interactions. 
Although nonlinear Hamiltonian $\hat{\mathcal{H}}_{\rm int}$ is invariant under time reversal, it admits existence of the asymmetric states different from Eq.~\eqref{TREigenStates}.
Recently, a striking example of SSB in the persistent polariton currents subject to the TE-TM splitting was predicted in \cite{Sedov2021SciRep}. 
In particular, the symmetry breaking was manifested in the excitation  of co-rotating circularly polarized polariton currents with almost equal occupancies.
In the considered system, the effect of the TE-TM splitting was rather strong and dominate the kinetic energy of the circular polariton motion which corresponds to the limit $\Delta \gg 1$ in the notations adopted in our paper. 
In contrast, we focus on the opposite limit of weak SOC \cite{yulin2022}, namely at $\Delta < 1$. 
This regime allows us to approach the problem analytically and reveal the family of distinct symmetry-breaking scenarios. 

Since in the ring geometry, the TE-TM splitting strength is inversely proportional to the squared ring width \cite{Kozin2018,Mukherjee2021} while the kinetic energy scales as $R^{-2}$, the regime $\Delta < 1$ can be achieved in the wide rings with small radius.

\subsection{Symmetry breaking in the scalar case}

We start by reviewing SSB phenomenon in the spinless case which is realised at vanishing TE-TM splitting, $\Delta=0$. 
The potential embedded in the ring breaks rotational symmetry and prohibits circulation of the single-particle states \cite{Demirchyan2022}. 
The self-interactions allow for breaking this symmetry and trigger polariton rotation either in the CW or CCW directions.

The properties of the broken-symmetry states can be revealed either numerically \cite{Demirchyan2022} or analytically in some specific cases \cite{Seaman2005,Bronski2001}.
We search for an approximated steady-state solutions representing condensate state as a combination of two contra-propagating waves with $+2 \pi$ and $-2\pi$ phase winding:
\begin{equation}\label{ScalarAnsatz}
    \Psi(\phi) = A e^{i\phi} + B e^{-i\phi}.
\end{equation}
This approximation works best with the shallow harmonic potential $U= 2V_0 \cos(2\phi)$  at $V_0 \ll 1$.  
Substituting ansatz \eqref{ScalarAnsatz} into the scalar analog of Eq.~\eqref{GPE} and neglecting the contribution of the higher order angular harmonics one obtains the system analogous to the nonlinear dimer model \cite{Jensen1982}:
\begin{subequations}\label{TwoModeModel}
\begin{eqnarray}
    \left(\mu-1\right) A &=& V_0 B + g\left(\left|A\right|^2 + 2\left|B\right|^2 \right)A,\\
    \left(\mu-1\right) B &=& V_0 A + g\left(\left|B\right|^2 + 2\left|A\right|^2 \right)B.
\end{eqnarray}
\end{subequations} 
The symmetric solutions of Eqs.~\eqref{TwoModeModel} respect parity, $A = \pm B$ and have zero OAM, $\ell = (\left|A\right|^2 - \left|B\right|^2)/\rho = 0$.
The asymmetric states with $\left|A\right|^2 = \rho/2\left(1\pm\sqrt{1-\rho_{\rm c}^2/\rho^2}\right)$ and $\left|B\right|^2 = \rho/2\left(1\mp\sqrt{1-\rho_{\rm c}^2/\rho^2}\right)$  appear above the critical polariton density
\begin{equation}\label{CriticalV0}
    \rho_{\rm c} = 2 V_0/g.
\end{equation}
The fingerprint of symmetry-broken states is a nonzero OAM $\ell = \pm \sqrt{1-\rho_{\rm c}^2/\rho^2 }$, which appears at $\rho > \rho_{\rm c}$. Far away from the bifurcation point $\rho \gg \rho_{\rm c}$, the asymmetric solutions are close to the left- and right-hand vortices when all polaritons is accumulated either in the CW, $\Psi\approx A e^{i\phi}$, or in the CCW wave, $\Psi\approx B e^{-i\phi}$, respectively.

\subsection{Symmetry breaking in the spinor polariton condensate}

In the spinor case, the criterion of the symmetry breaking has to be revised.
At the finite strength of the TE-TM splitting, the symmetry arguments ensure formation of the linear polarized states with vanishing total OAM. 
Therefore, we characterize the SSB phenomenon using two parameters.
Namely, we expect that symmetry breaking  either triggers global rotation of the condensate, $\ell^{\rm tot} \neq 0$, or induces imbalance between polarization components which can be quantified by the average degree of circular polarization (DCP):
\begin{equation}
    P_c= {N_{\rm pol}}^{-1} \int \left( |\Psi_+|^2 - |\Psi_-|^2\right) d \phi.
\end{equation}

First, we address the problem numerically. 
In particular, we search for the stationary solutions of Eq.~\eqref{GPE} iteratively using average polariton density $\rho$ as a scanning parameter.
As in the linear case, we consider a point-like defect potential $U = 2\pi V_0\delta(\phi)$.

\begin{figure}
    \centering
    \includegraphics[width = \linewidth]{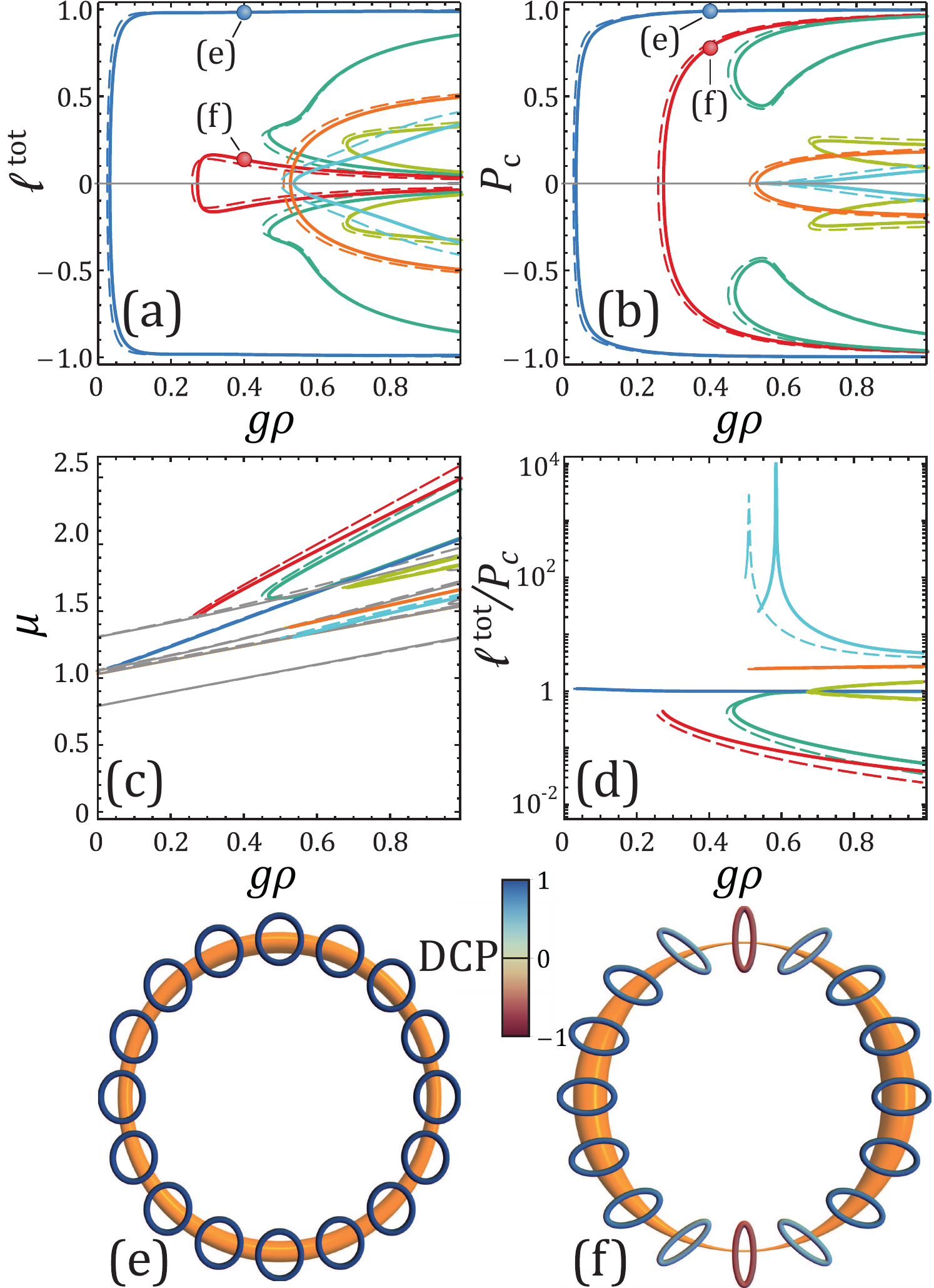}
    \caption{Symmetry-broken states of the ring-shaped spinor polariton condensate with broken rotational symmetry. (a) The total OAM per particle $\ell^{\rm tot}$ vs average polariton density $\rho$ at $V_0=0.05$ and $\Delta=0.25$. The color lines correspond to the states with broken symmetry, while $\hat{\mathcal{T}}$-symmetric states with $\ell^{\rm tot}=0$ are shown in grey. Solid lines correspond to the steady-state solutions of the full problem \eqref{GPE}. The dashed curves are obtained from the simplified four-wave model \eqref{FourModeModel}. (b) The $P_c(\rho)$-dependence and (c) the $\mu(\rho)$-dependence of the solutions shown in (a). (d) The ratio $\left| \ell^{\rm tot}/P_c\right|$ for the states with broken symmetry.
    The scale of the horizontal axes in (a)-(d) corresponds to the average blue shift $g\rho$ of the homogeneous circularly polarized polariton condensate.
    (e) Polarization pattern of the symmetry-broken state corresponding to the circularly polarized persistent current [shown in blue in (a)-(d)] at  $g \rho = 0.4$.  Polarization ellipses are shown in color indicating the local degree of circular polarization. The thickness of the orange ring corresponds to the local polariton density $\left|\Psi_+\right|^2 + \left|\Psi_-\right|^2$. (f) The same as in (e) but for the circularly polarized state with weak total current [shown in red in (a)-(d)].}
    \label{fig2}
\end{figure}

Figures~\ref{fig2}a--c demonstrate  the typical example of the bifurcation diagrams on the  $(\rho,\ell^{\rm tot})$,  $(\rho,P_c)$ and $(\rho,\mu)$ parameter planes. 
Note that the linear solutions discussed in Sec.~\ref{SecSPstates} retain their $\hat{\mathcal{T}}$-symmetric properties in the nonlinear regime, see the grey lines. 
Besides them, a plethora of symmetry-broken states with $\ell^{\rm tot} \neq 0$ or $P_c \neq 0$ emerge at large polariton densities.

\begin{figure}
    \centering
    \includegraphics[width = \linewidth]{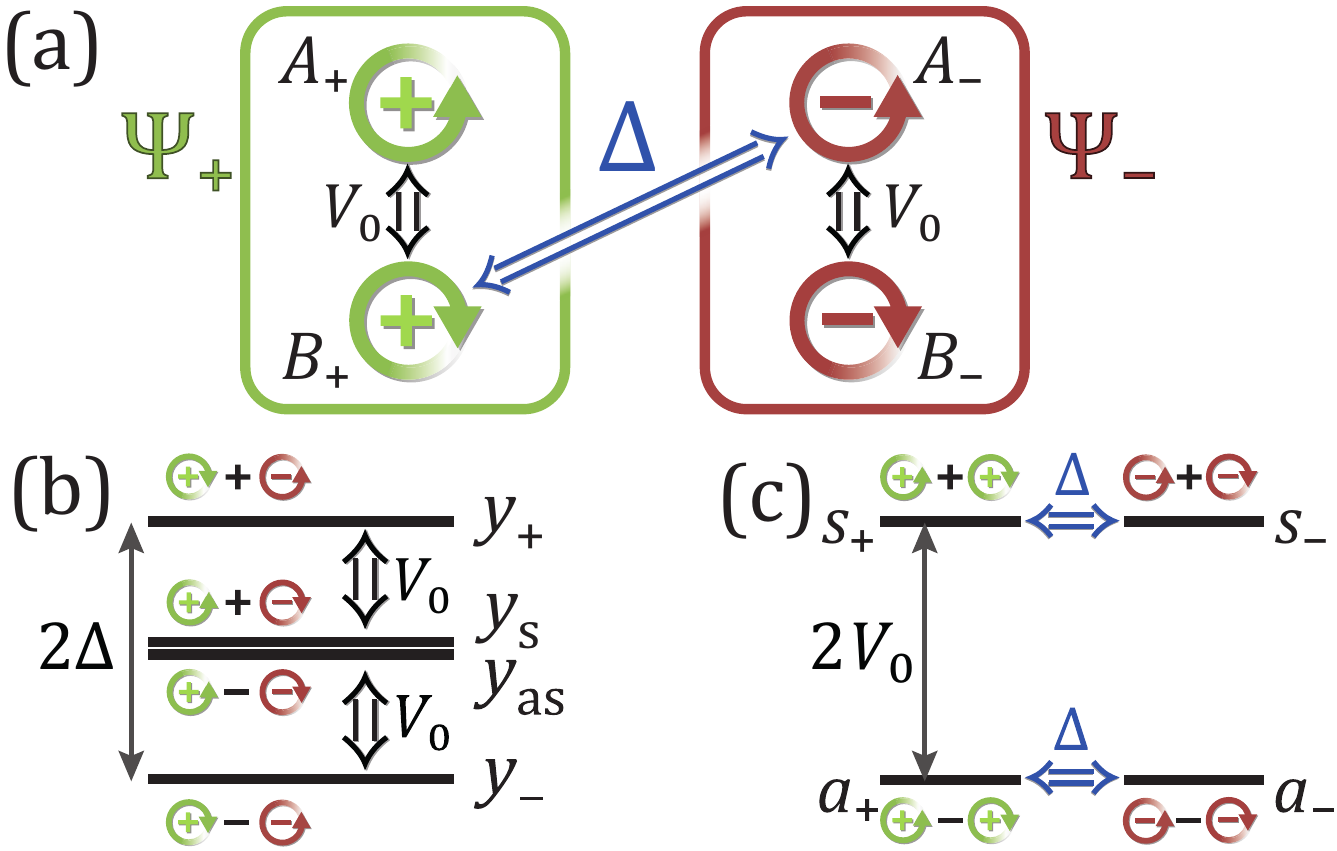}
    \caption{The four-wave model of the symmetry breaking in a spinor polariton ring. 
    (a) The level diagram in the four-wave approximation. The TE-TM splitting couples contra-propagating states with opposite circular polarizations, while the potential mixes CW and CCW waves within the same polarization. 
    (b) The simplified energy diagram in the regime of the dominating TE-TM splitting, $\Delta \gg V_0$. The structure of the linear eigenstates $y_{\rm s, as}$ and $y_{\pm}$ are explained schematically with the circular-arrow badges near the level bars. 
    (c) The simplified energy diagram at the weak TE-TM splitting, $\Delta \ll V_0$. The linear eigenstates correspond to the symmetric $s_{\pm}$ and antisymmetric $a_{\pm}$ circularly polarized states. }
    \label{fig3}
\end{figure}

Figure~\ref{fig2}d demonstrates the ratio between the OAM and DCP, $\ell^{\rm tot}/P_c$, for the states with broken symmetry which allows to classify them into three groups. 
The first one contains the persistent current states with vanishing DCP, $\ell^{\rm tot} \gg P_c$ (the cyan and the orange curves). 
In the opposite side, there are states with the dominating circular polarization and vanishing net circulation, $\ell^{\rm tot} \ll P_c$ (the red and the green curves). 
In between, there are those states which combine nonzero persistent current and finite circular polarization, $\ell^{\rm tot} / P_c \approx 1$, see the blue curve in Fig.~\ref{fig3}.

To gain more insights into the revealed SSB scenarios, we again resort to the simplified model. 
In particular, we project Eq.~\eqref{GPE} into the subspace consisted of four states: two opposite vortices with $\pm 2\pi$-phase winding 
in each polarization:
\begin{equation}\label{SpinorAnsatz}
    \Psi_{\pm}(\phi) = A_{\pm} e^{i\phi} + B_{\pm} e^{-i\phi}.
\end{equation}
These basis states correspond to the degenerate quadruplet of the  linear eigenstates $\{\mu_{0-},\mu_{2-},\mu_{-2-},\mu_{0+}\}$ at $\Delta=V_0=0$, see Fig.~\ref{fig1}a. 
The spin-less potential couples counter-rotating waves within the same circular polarization, while the splitting of TE and TM modes mixes states with $\ell^{-} - \ell^{+}= 2$ \cite{Flayac2010,Sedov2021SciRep}. 
Within the given subspace, these are $A_{-}$ and $B_{+}$ modes. The schematic structure of the coupled basis levels is show in Fig.~\ref{fig3}a.
The interaction with other angular harmonics can be safely neglected provided their detuning from the $\ell^{\pm}=\pm 1$ states exceeds the strength of the linear coupling from the SOC and the potential. 
It requires  $\Delta < 1$ and $V_0 < 1$. 
In this limit, the projection procedure yields:
\begin{subequations}\label{FourModeModel}
\begin{eqnarray}
    \bar{\mu} A_{+} &=& V_0 B_{+} + g\left(\left|A_{+}\right|^2 + 2\left|B_{+}\right|^2 \right) A_{+},\phantom{aaaaaaaaaa}\\
    \bar{\mu} B_{+} &=& V_0 A_{+} + g\left(\left|B_{+}\right|^2 + 2\left|A_{+}\right|^2 \right) B_{+} + \Delta A_{-},\\
    \bar{\mu} A_{-} &=& V_0 B_{-} + g\left(\left|A_{-}\right|^2 + 2\left|B_{-}\right|^2 \right) A_{-} + \Delta B_{+},\\
    \bar{\mu} B_{-} &=& V_0 A_{-} + g\left(\left|B_{-}\right|^2 + 2\left|A_{-}\right|^2 \right) B_{-},
\end{eqnarray}
\end{subequations} 
where $\bar{\mu} = \mu-1-V_0$. 
Within the specified parameter range, this simplified model demonstrates a good coincidence with the full GPE \eqref{GPE}, see the solid and the dashed lines in Fig.~\ref{fig2}. 
In general, Eqs.~\eqref{FourModeModel} have to be solved numerically.
However, in some specific cases the analytical predictions can be made. 
In particular, we consider two limits which allow for prediction of the threshold polariton density corresponding to the formation of the broken symmetry states.

\subsection{The limit of the dominating SOC $\Delta \gg V_0 $}
\begin{figure}
    \centering
    \includegraphics[width = \linewidth]{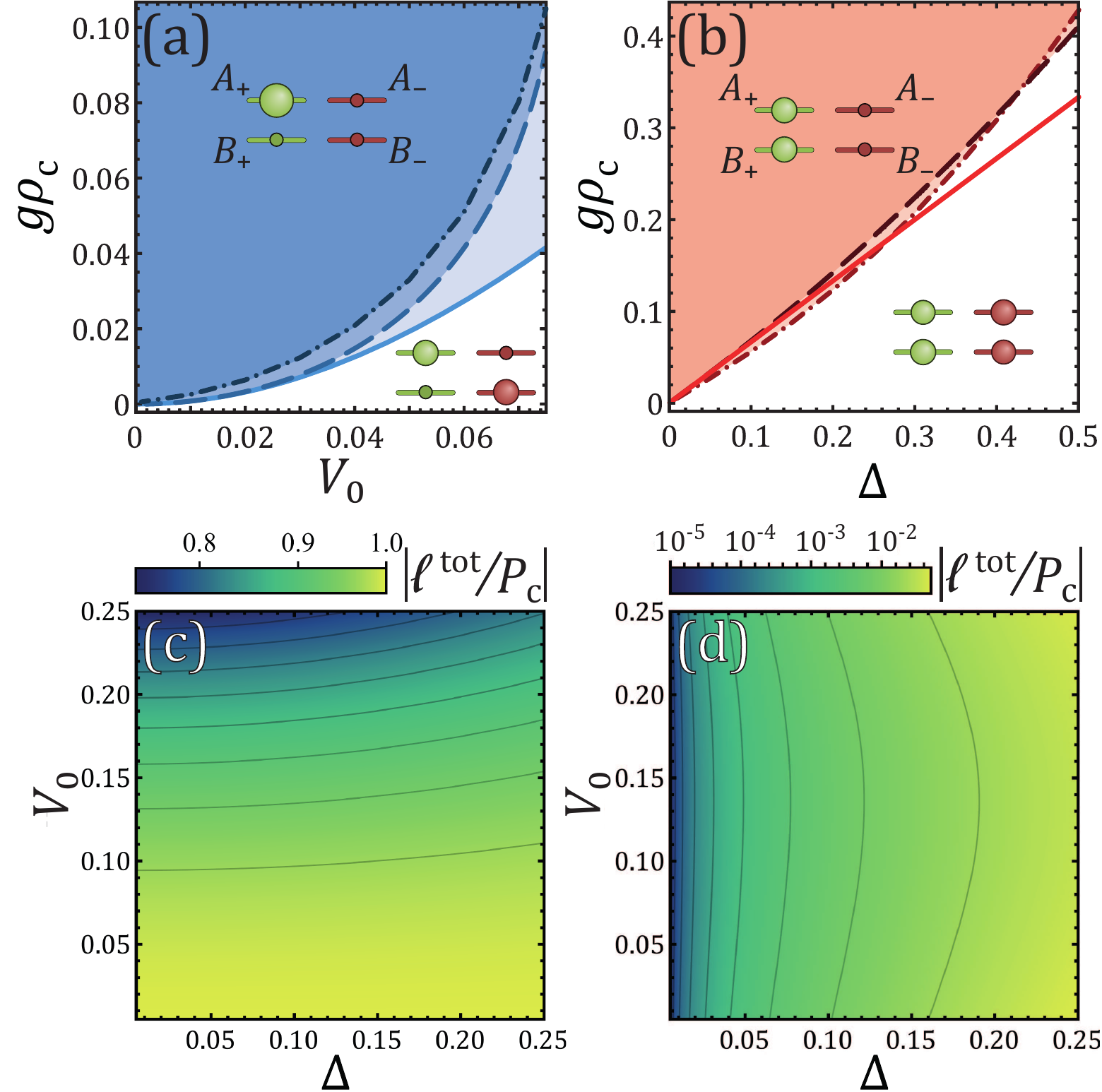}
    \caption{(a) The threshold polariton density above which the circularly polarized persistent current shown in Fig.~\ref{fig2}e appears. The solid line corresponds to the analytical estimation \eqref{Eq:StrongSocThreshold} while the dash-dotted line was obtained from the full GPE \eqref{GPE} with the delta-defect barrier $U(\phi) = 2\pi V_0\delta(\phi)$ at $\Delta = 0.25$. The dashed line demonstrates predictions of the four-wave model \eqref{FourModeModel}.  
    The level diagram inside the shaded (white) region schematically demonstrates occupancy of the four vortex states shown in Fig.~\ref{fig3}a in the symmetry-broken (symmetric) state. 
    (b) The threshold polariton density corresponding to the bifurcation of the circularly polarized state with the vanishing circulation shown in Fig.~\ref{fig2}f at the defect strength $V_0=0.25$. The solid line corresponds to Eq.~\eqref{Eq:WeakSocThreshold}, the dashed and the dash-dotted curves have the same meaning as in (a).
    (c) and (d) The dependencies of the $\left|\ell^{\rm tot}/P_c\right|$-parameter at $g\rho = 0.75$ as functions of $\Delta$ and $V_0$  for the states described in (a) and (b), respectively.}
    \label{fig4}
\end{figure}

First, we consider the state which bifurcates in a pitchfork event at the small polariton density, -- see the blue curve in Figs.~\ref{fig2}a,b. 
At $V_0 = 0$ the structure of the linear eigenmodes are governed by the TE-TM splitting only. 
According to Eqs.~\eqref{FourModeModel}, two states $y_\pm = \left(A_- \pm B_+\right)/\sqrt{2}$ are split by $2\Delta$ while the uncoupled states $A_+$ and $B_-$ remain degenerate. 
At finite $V_0 \ll \Delta$, the state $y_\pm$ primarily couples to the symmetric (antisymmetric) combination of the degenerate modes $y_{\rm s, as} = \left(A_+ \pm B_-\right)/\sqrt{2}$. 
As a result, the energy of the $y_{\rm s}$-state decreases by $\delta = \left( \sqrt{\Delta^2 + 4V_0^2 } - \Delta\right)/2$ while the $y_{\rm as}$-state shifts upward by $\delta$, see the schematic energy diagram in Fig.~\ref{fig3}b. 
Herewith, separating the $y_{\rm s, as}$-doublet from Eqs.~\eqref{FourModeModel} by neglecting nonlinear mixing with the $y_\pm$-states and turning back to the basis of $A_+$ and $B_-$ waves, 
one obtains
\begin{subequations} \label{Eq:StrongSocLimit}
    \begin{eqnarray}
   \bar{\mu} A_+ &=& \delta B_- + g \left|A_+ \right|^2 A_+,\\
   \bar{\mu} B_- &=& \delta A_+ + g \left|B_- \right|^2 B_-.    
    \end{eqnarray}
\end{subequations}
This system is analogous to the two-mode model \eqref{TwoModeModel}. The \textit{internal} parity symmetry governed by the $\delta$-parameter is established between the waves decoupled from the TE-TM splitting, see Fig.~\ref{fig3}a.
The nonlinear symmetry-breaking solutions of Eqs.~\eqref{Eq:StrongSocLimit} exist above the critical density
\begin{equation} \label{Eq:StrongSocThreshold}
    \rho_{\rm c} = \left( \sqrt{\Delta^2 +4V_0^2 } - \Delta\right)\left/ g\right..
\end{equation}
The $V_0$-dependence of the SSB threshold density obtained numerically from the four-wave model (dashed line) and from the full GPE (dash-dotted line) is shown in Fig.~\ref{fig4}a. 
The analytical estimation \eqref{Eq:StrongSocThreshold} demonstrates a good coincidence with the numerical results only  at $V_0 \ll 1$. 
Beyond this limit, the truncation to the two mode-model fails. 
However, the four-wave model \eqref{FourModeModel} describes the onset of the SSB phenomenon qualitatively well.

Away from the SSB bifurcation point, $\rho > \rho_{\rm c}$, polariton density is accumulated either in $A_+$ or $B_-$ vortex, see the schematic energy diagram in the inset to Fig.~\ref{fig4}a. 
Therefore, these states possess large DCP and almost unit OAM simultaneously. 
An example of the corresponding polarization pattern is shown in Fig.~\ref{fig2}e. 
Figure~\ref{fig4}c demonstrates the variation of the ratio $\left|\ell^{\rm tot}/P_c \right|$ in the parameter plane $(\Delta,V_0)$ at the fixed polariton density. 
This ratio remains close to unity which means that the symmetry-broken states retain their properties in a wide range of parameters.

\subsection{The limit of weak SOC $\Delta \ll V_0$}

A similar truncation procedure can be carried out in the opposite limit $\Delta \ll V_0$  which implies that the scattering on the potential defect dominates the wave mixing due to the TE-TM splitting.
In particular, at the vanishing TE-TM splitting, $\Delta=0$, the linear eigenstates of Eqs.~\eqref{FourModeModel} correspond to the symmetric $s_{\pm} = \left( A_{\pm} + B_{\pm}\right)/\sqrt{2}$ and antisymmetric $a_{\pm} = \left( A_{\pm} - B_{\pm}\right)/\sqrt{2}$ circularly polarized states which are split in energy by $2V_0$ and two-fold degenerate in polarization. 
The simplified scheme of the energy levels is shown in Fig.~\ref{fig3}c. 
The weak SOC $\Delta \ll V_0$ primarily mixes degenerate states while the coupling between the detuned $s_{\pm}$-states and $a_{\mp}$-states remains negligibly weak. 
Herewith, rewriting Eqs.~\eqref{FourModeModel} in the basis of $s_{\pm}$ and $a_{\pm}$ and leaving the resonant terms only we split the system into a pair of independent nonlinear {dimers}:
\begin{subequations}\label{Eq:WeakSocLimit}
\begin{eqnarray}
    \left(\bar{\mu} - V_0 \right) s_\pm &=&   \frac{3g}{2} \left|s_\pm\right|^2 s_\pm + \frac{\Delta}{2} s_\mp, \\
    \left(\bar{\mu} + V_0 \right) a_\pm &=&   \frac{3g}{2} \left|a_\pm\right|^2 a_\pm - \frac{\Delta}{2} a_\mp.
\end{eqnarray}
\end{subequations} 

Both symmetric and antisymmetric dimers admit existence of the asymmetric solutions which break the parity symmetry imposed by the weak coupling from the TE-TM splitting.
In particular, two pairs of asymmetric states with $\left|s_+\right| \neq \left|s_-\right|$ and $\left|a_+\right| \neq \left|a_-\right|$ bifurcate from the $s_\pm$ and  $a_\pm$  doublets above the critical dimensionless polariton density given by
\begin{equation}\label{Eq:WeakSocThreshold}
    \rho_{\rm c} = \frac{2\Delta}{3g}.
\end{equation}
This analytical estimation of the threshold polariton density agrees well with the predictions of both simplified \eqref{FourModeModel} and full GPE models, see Fig.~\ref{fig4}b. Note that the coincidence is good even beyond the considered limit of weak SOC, $V_0 \gg \Delta$.

The SSB event disrupts the balance between polarization components. 
However the chiral symmetry between CW and CCW currents of the same polarization gets only weakly broken since the intra-polarization coupling  dominates at $V_0 \gg \Delta$, see the inset in Fig.~\ref{fig4}b.  
Therefore, the new symmetry breaking states are characterised by the non-zero DCP and weak total current $\ell^{\rm tot} \approx 0$. 
These states are shown with red lines in Fig.~\ref{fig2}a--d. 
The example of the corresponding polarization pattern is demonstrated in Fig.~\ref{fig2}f. 
Note that these properties are retained even beyond the limit $\Delta \gg V_0$  as it is shown on the map of the $\left|\ell^{\rm tot}/P_c \right|$-ratio on the parameter plane $(\Delta,V_0)$, see Fig.~\ref{fig4}d.

\section{Conclusions}
Nonlinear effects that lead to breaking of the symmetry between contra propagating waves in the split-ring geometry were previously studied in the spinless systems such as bosonic condensates and whispering-gallery microcavities. The present manuscript systematically extends these concepts into the spinor case accounting for the mechanism of the spin to optical angular momentum coupling associated with the effect of the TE-TM splitting that is characteristic of semiconductor microcavities. Our findings pave the way for the creation and manipulation of the spin-polarized persistent currents of exciton polaritons. 

We specifically focused on the split-ring geometry, where circular currents are suppressed in the case of a scalar polariton condensate in the linear regime \cite{Demirchyan2022}. We have shown that in the linear and weakly nonlinear regimes, the spinor condensate structure is governed by the underlying symmetry of the effective spin-orbit coupling mechanism. This symmetry implies invariance under simultaneous inversion of the rotation direction and swapping of circular polarizations.
In this case, the condensate is linearly polarized and has no net circulation while  circularly polarized polaritons rotate in opposite directions.

In the nonlinear regime, the many-body states with reduced symmetry appear spontaneously. 
We reveal several types of the broken-symmetry states including circularly polarized persistent currents, circularly polarized condensate with no rotation and the persistent current state with the small degree of circular polarization. 
The properties of the states of the first and the second types can be described with the use of the simplified analytical model. 
In particular, we predict the values of the critical polariton density corresponding to the symmetry breaking threshold for these states.  

\begin{acknowledgments}
This work was supported by the Russian Science Foundation grant No. 22-72-00061.
A.K. acknowledges the Spin Optics
Laboratory at Saint-Petersburg State University (Grant No.
94030557) and the Westlake
University, Project No. 041020100118 and Program No.
2018R01002 funded by Leading Innovative and Entrepreneur
Team Introduction Program of Zhejiang Province of China.
\end{acknowledgments}

%

\end{document}